\documentclass[twocolumn,superscriptaddress,nobibnotes,footinbib,floatfix,aps,prb,citeautoscript,reprint]{revtex4-2}

\usepackage{graphicx}
\usepackage{epsfig}
\usepackage{subfigure}
\usepackage{color}
\usepackage{latexsym}
\usepackage{amsmath,bm,multirow}
\usepackage{amsfonts}
\usepackage{dcolumn}           
\usepackage{natbib}
\usepackage[colorlinks,linkcolor=blue,citecolor=green]{hyperref}
\usepackage{physics}

\usepackage[draft]{changes} 
\usepackage[normalem]{ulem}
\definechangesauthor[name={yi}, color=blue]{yi}

\newcommand{\portland}{Department of Mechanical and Materials Engineering, Portland State University, Portland, OR 97201, USA}

\usepackage[draft]{changes} 
\usepackage[normalem]{ulem}
\definechangesauthor[name={Maria}, color=red]{Maria}
\definechangesauthor[name={Yi}, color=blue]{Yi}
\usepackage{soul}
\begin{document}

\title{Quintic-Anharmonicity-Assisted Three-Phonon Scattering: \\ A Previously Overlooked Same-Order Channel to Four-Phonon Scattering}

\author{Yi Xia}
\email{yimaverickxia@gmail.com; yxia@pdx.edu}
\affiliation{\portland}

\date{\today}

\begin{abstract}

Four-phonon scattering is widely regarded as the leading higher-order anharmonic correction to conventional three-phonon description of anharmonic scattering and lattice thermal transport. Here we show that diagrammatic perturbation theory contains another phonon linewidth contribution of the same perturbation order, arising from the coupling between cubic and quintic anharmonic vertices. We derive and implement this contribution from first principles and identify it as a quintic-anharmonicity-assisted three-phonon scattering channel. Although this process shares energy-conservation structure of the conventional three-phonon process, its perturbative order and temperature dependence resemble those of the conventional four-phonon process. For Si, we find that phonon scattering from this additional channel  is comparable to four-phonon scattering over a broad frequency and temperature range, leading to a similar accelerated reduction of lattice thermal conductivity at high temperatures. We further show that in strongly anharmonic AgCl, this channel can become comparable to the ordinary three-phonon scattering even at room temperature. These results demonstrate that the commonly observed high-temperature enhanced scattering and suppression of lattice thermal conductivity might not be attributed uniquely to four-phonon process, and establish quintic-anharmonicity-assisted three-phonon scattering as a previously overlooked same-order channel in anharmonic lattice dynamics, which may be leveraged to uncover hidden microscopic thermal transport mechanisms.
\end{abstract}

\maketitle


\textit{Introduction.} Phonons play an important role in a wide range of materials properties, including lattice thermal transport~\cite{nellis2008heat,LINDSAY2018106}, structural stability~\cite{Cowley1968,SSCHA}, carrier relaxation~\cite{giustino2017electron, bernardi2016first}, and optical response~\cite{antonius2022theory} when lattice vibrations are coupled to electrons and photons, respectively. In crystalline solids, the conventional description of lattice dynamics and phonon transport is based on weakly interacting phonon quasiparticles, whose lifetimes are primarily limited by anharmonic scattering~\cite{wallace1998thermodynamics,ziman1960electrons}. At moderate temperatures, the dominant intrinsic scattering mechanism is usually the three-phonon (3ph) processes arising from cubic anharmonicity~\cite{Cowley1968,srivastava1990physics}. This picture leads to the familiar temperature dependence of lattice thermal conductivity ($\kappa_{l}$) that approximately follows $T^{-1}$~\cite{LINDSAY2018106,mcgaughey2025phonon}. 

Recent first-principles studies have shown~\cite{Tianli2016,Tianli2017,pbte2018,ravichandran2018unified1}, however, that this conventional 3ph picture is often incomplete. In particular, four-phonon (4ph) processes arising from quartic anharmonicity provide an additional phonon scattering channel, especially at elevated temperatures~\cite{Tianli2017,ravichandran2020phonon} and in strongly anharmonic materials~\cite{rczbprx,xia2020microscopic}. The inclusion of 4ph scattering has been shown to enhance the scattering rates, reduce the predicted $\kappa_{l}$, improve agreement with experiment in several materials, and produce a stronger than $T^{-1}$ temperature dependence in $\kappa_{l}$~\cite{Tianli2017,yang2019stronger,rczbprx,ravichandran2020phonon}. As a result, 4ph scattering is now widely regarded as the leading higher-order correction beyond ordinary 3ph picture~\cite{McGaughey2019}.

Recognizing the importance of 4ph scattering, it is instructive to systematically examine the hierarchy of scattering processes arising from higher-order anharmonicity. In the established quantum-field-theoretical formulation of lattice~\cite{maradudin1962scattering, Cowley1968,tripathi1974self}, phonon linewidths are obtained from the imaginary part of the phonon self-energy. Different anharmonic scattering channels are represented by distinct Feynman diagrams constructed from anharmonic vertices~\cite{maradudin1962scattering,mahan2000many}, which correspond to force constants expressed in normal-mode coordinates. The relative order of these diagrams can be classified using the Van Hove's ordering parameter $\lambda$~\cite{van1961problems}, defined as the ratio between a characteristic atomic displacement and the interatomic distance. Under this ordering, an $n$th-order anharmonic vertex scales as $V_n \sim O(\lambda^{n-2})$~\cite{tripathi1974self}. Consequently, the conventional 3ph linewidth contribution from two cubic vertices scales as $V_3V_3 \sim O(\lambda^2)$, whereas the conventional 4ph contribution from two quartic vertices scales as $V_4V_4 \sim O(\lambda^4)$. Importantly, this same order also contains a mixed cubic-quintic contribution, $V_3V_5 \sim O(\lambda^4)$, which contributes to the imaginary part of the phonon self-energy. Thus, from the viewpoint of perturbation order, this mixed $V_3V_5$ channel should be placed on the same footing as the conventional 4ph process.

Despite this simple ordering argument, the $V_3V_5$ contribution has not, to the best of our knowledge, been explicitly evaluated. This omission is understandable because the process requires quintic anharmonic force constants, whose first-principles evaluation and application to lattice dynamics and phonon transport have only recently been demonstrated in the context of five- and six-phonon scatterings~\cite{xia2025first}. In this Letter, we identify and numerically evaluate this previously overlooked scattering channel, which we term quintic-anharmonicity-assisted three-phonon ($V_3V_5$-3ph) scattering. We implement the $V_3V_5$-3ph linewidth contribution from first principles and compare it directly with the conventional $V_3V_3$-3ph and $V_4V_4$-4ph scattering. Using Si as a benchmark system, we show that the $V_3V_5$-3ph scattering rates are comparable in magnitude to $V_4V_4$-4ph scattering over a broad range of phonon frequencies and temperatures. When included in the calculation of $\kappa_{l}$, this channel produces a reduction and temperature dependence similar to those caused by the conventional $V_4V_4$-4ph scattering. We further examine AgCl as a more strongly anharmonic material~\cite{rczbprx} and find that the relative importance of the $V_3V_5$-3ph channel can be substantially enhanced, becoming comparable to ordinary $V_3V_3$-3ph scattering near room temperature.

\begin{figure*}[htp]
\includegraphics[width = 1.0\linewidth]{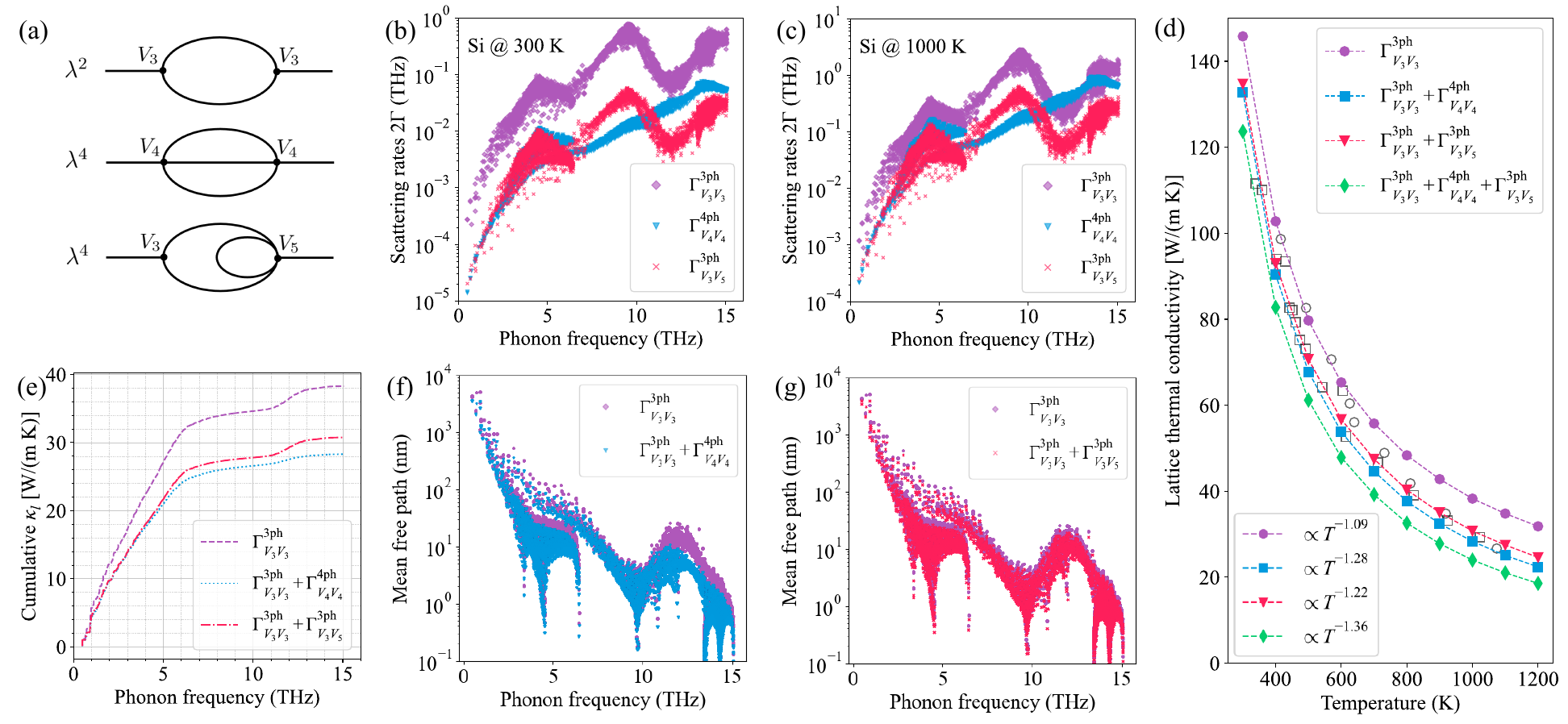}
	\caption{
    (a) Diagrammatic representation of the three scattering channels considered (from top to bottom): the conventional lowest-order three-phonon (3ph) process from two cubic anharmonic vertices $V_3V_3$, the conventional four-phonon (4ph) process from two quartic vertices $V_4V_4$, and the newly added quintic-anharmonicity-assisted 3ph process involving coupled cubic and quintic vertices $V_3V_5$. $\lambda$ is the Van Hove's ordering parameter, indicating the magnitude of perturbation order.
    (b,c) Mode-resolved scattering rates ($2\Gamma$) of Si as a function of phonon frequency at 300 K and 1000 K, respectively, comparing $\Gamma^{\mathrm{3ph}}_{V_3V_3}$, $\Gamma^{\mathrm{4ph}}_{V_4V_4}$, and $\Gamma^{\mathrm{3ph}}_{V_3V_5}$.
    (d) Temperature-dependent lattice thermal conductivity ($\kappa_{l}$) of Si accounting for the hierarchy of various phonon-scattering channels, i.e., $\Gamma^{\mathrm{3ph}}_{V_3V_3}$, $\Gamma^{\mathrm{3ph}}_{V_3V_3}+\Gamma^{\mathrm{4ph}}_{V_4V_4}$, $\Gamma^{\mathrm{3ph}}_{V_3V_3}+\Gamma^{\mathrm{3ph}}_{V_3V_5}$, and $\Gamma^{\mathrm{3ph}}_{V_3V_3}+\Gamma^{\mathrm{4ph}}_{V_4V_4}+\Gamma^{\mathrm{3ph}}_{V_3V_5}$.
    The corresponding power-law temperature dependences are indicated as $T^{-1.09}$, $T^{-1.28}$, $T^{-1.22}$, and $T^{-1.36}$, respectively.
    Empty circles~\cite{glassbrenner1964thermal} and squares~\cite{abeles1962thermal} denote experimental measurements.
    (e) Cumulative $\kappa_{l}$ obtained as a function of phonon frequency in Si at 1000~K, showing the reduction of $\kappa_{l}$ when $\Gamma^{\mathrm{4ph}}_{V_4V_4}$ or $\Gamma^{\mathrm{3ph}}_{V_3V_5}$ is added to $\Gamma^{\mathrm{3ph}}_{V_3V_3}$.
    (f,g) Mode-resolved phonon mean free paths calculated from  $\Gamma^{\mathrm{3ph}}_{V_3V_3}$ alone and after including either $\Gamma^{\mathrm{4ph}}_{V_4V_4}$ or $\Gamma^{\mathrm{3ph}}_{V_3V_5}$ in Si at 1000~K.
}
	\label{fig:diagram}
\end{figure*}



\textit{Methods.} 
We evaluate anharmonic phonon scattering within the diagrammatic perturbation theory of lattice dynamics based on the pioneering work by Maradudin and Fein~\cite{maradudin1962scattering}, Cowley~\cite{Cowley1968}, and Tripathi and Pathak~\cite{tripathi1974self}, amongst others~\cite{procacci1992anharmonic}. This approach enables a systematic evaluation of multiphonon scattering processes through the calculation of phonon self-energies, with the corresponding Feynman diagrams analyzed according to established diagrammatic rules~\cite{maradudin1962scattering}. Moreover, these diagrams can be classified by the perturbation order using the Van Hove's ordering parameter $\lambda$~\cite{van1961problems}. Specifically, as shown by Tripathi and Pathak~\cite{tripathi1974self}, there are two diagrams of $O(\lambda^2)$ and five diagrams of $O(\lambda^4)$ within up to the second-order perturbation theory, respectively (see Fig.~1 and Fig.~2 of Ref.[\onlinecite{tripathi1974self}]). Among these seven diagrams, only three diagrams contribute to the imaginary part of phonon self-energy, while the others merely contribute to phonon frequency shift. These three diagrams, shown in Fig.~\ref{fig:diagram}(a), are the focus of this study. From top to bottom, they correspond to the conventional three-phonon process from $V_3V_3$, the conventional four-phonon process from $V_4V_4$, and the mixed cubic--quintic process from $V_3V_5$. Following the standard diagrammatic rules for anharmonic phonons and using Green’s function techniques~\cite{maradudin1962scattering,tripathi1974self,mahan2000many,xia2025first}, the corresponding linewidths ($\Gamma$) for a phonon mode with wave vector $\mathbf{q}$ and branch index $j$ can be derived as follows:

\begin{widetext}
\begin{align}
\Gamma^{3\mathrm{ph}}_{V_3V_3}(\mathbf{q}j)
=& \frac{18}{\hbar^2}
\sum_{\mathbf{q}_1j_1}\sum_{\mathbf{q}_2j_2}
V_3(\mathbf{q}j;\mathbf{q}_1j_1;\mathbf{q}_2j_2)
V_3(-\mathbf{q}j;-\mathbf{q}_1j_1;-\mathbf{q}_2j_2)
\Delta(\mathbf{q}+\mathbf{q}_1+\mathbf{q}_2)
\operatorname{Im}\mathcal{F}(\omega+i\epsilon,\omega_1,\omega_2),
\label{eq:g33}
\\
\Gamma^{4\mathrm{ph}}_{V_4V_4}(\mathbf{q}j)
=&\frac{96}{\hbar^2}
\sum_{\mathbf{q}_1j_1}\sum_{\mathbf{q}_2j_2}\sum_{\mathbf{q}_3j_3}
V_4(\mathbf{q}j;\mathbf{q}_1j_1;\mathbf{q}_2j_2;\mathbf{q}_3j_3)
V_4(-\mathbf{q}j;-\mathbf{q}_1j_1;-\mathbf{q}_2j_2;-\mathbf{q}_3j_3)
\nonumber \\
&\times
\Delta(\mathbf{q}+\mathbf{q}_1+\mathbf{q}_2+\mathbf{q}_3)
\operatorname{Im}
\mathcal{F}(\omega+i\epsilon,\omega_1,\omega_2,\omega_3),
\label{eq:g44}
\\
\Gamma^{3\mathrm{ph}}_{V_3V_5}(\mathbf{q}j)
=&\frac{360}{\hbar^2}
\sum_{\mathbf{q}_1j_1}\sum_{\mathbf{q}_2j_2}\sum_{\mathbf{q}_3j_3}
V_3(\mathbf{q}j;\mathbf{q}_1j_1;\mathbf{q}_2j_2)
V_5(-\mathbf{q}j;-\mathbf{q}_1j_1;-\mathbf{q}_2j_2;\mathbf{q}_3j_3;-\mathbf{q}_3j_3)
\nonumber\\
&\times
\Delta(\mathbf{q}+\mathbf{q}_1+\mathbf{q}_2)
(2n_3+1)
\operatorname{Im}
\mathcal{F}(\omega+i\epsilon,\omega_1,\omega_2).
\label{eq:g35}
\end{align}
\end{widetext}

Here $V_n$ denotes the $n$th-order anharmonic vertex which represents the real-space force constants expressed in phonon normal coordinates, $\omega_i$ and $n_i$ are the phonon frequency and Bose-Einstein occupation of mode $(\mathbf{q}_i,j_i)$, respectively, and $\Delta$ enforces crystal-momentum conservation up to a reciprocal lattice vector. The function $\mathcal{F}$ contains the frequency denominators and thermal occupation factors generated by the corresponding diagram, and its imaginary part (Im$\mathcal{F}$) gives the energy-conservation delta functions in the limit $\epsilon= 0^+$. The total linewidth is related to the scattering rate (or reciprocal of phonon lifetime) by $1/\tau_{\mathbf{q}j}=2\Gamma_{\mathbf{q}j}$. The detailed expressions for $V_n$ and $\mathcal{F}$ can be found in the Supplementary Materials of Ref.[\onlinecite{xia2025first}].

Compared with Eqs.~\eqref{eq:g33} and \eqref{eq:g44}, which have already been widely studied in the literature~\cite{shengbte,tadano2014anharmonic,han2022fourphonon, barbalinardo2020efficient,nayeb2025thermacond}, Eq.~\eqref{eq:g35} highlights the distinct physical character of the mixed $V_3V_5$ scattering channel. Because $\Gamma^{3\mathrm{ph}}_{V_3V_5}$ contains the same $\operatorname{Im}\mathcal{F}(\omega+i\epsilon,\omega_1,\omega_2)$ factor and the same 3ph momentum-conservation condition as $\Gamma^{3\mathrm{ph}}_{V_3V_3}$, its frequency dependence is expected to resemble that of conventional 3ph scattering. However, the additional internally contracted phonon pair from the quintic vertex produces the factor $(2n_3+1)$, which enhances the temperature dependence. In the classical high-temperature limit, this additional Bose factor makes $\Gamma^{3\mathrm{ph}}_{V_3V_5}$ scale approximately as $T^2$, similar to $\Gamma^{4\mathrm{ph}}_{V_4V_4}$. Therefore, the $V_3V_5$-3ph process combines a 3ph-like phase space with a 4ph-like perturbative order and temperature dependence.

We implement this formalism and apply it to investigate anharmonic lattice dynamics and thermal transport in Si and AgCl. Silicon is used as a benchmark because 4ph scattering is known to become important at elevated temperatures~\cite{Tianli2017}, while higher-order five- and six-phonon scattering has been found to be relatively weak~\cite{xia2025first}. AgCl is examined as a more strongly anharmonic system with lower phonon frequencies~\cite{rczbprx,ouyang2023role}, where the thermally enhanced $V_3V_5$ channel is expected to become more prominent. Harmonic and anharmonic force constants are taken from our previous first-principles calculations~\cite{xia2025first,xia2025lattice} and are shown in Appendix~\ref{sec:ifcs}. All linewidths and $\kappa_{l}$ are evaluated using a $32\times32\times32$ phonon wave-vector mesh in the first Brillouin zone, accelerated by means of the subsampling~\cite{xia2025first} and stochastic sampling approaches~\cite{guo2024sampling}.

\textit{Results and discussion.} We first compare the mode-resolved scattering rates of the three channels in Si. Fig.~\ref{fig:diagram}(b) shows the scattering rates ($2\Gamma$) from the conventional 3ph scattering, $\Gamma^{3\mathrm{ph}}_{V_3V_3}$, the conventional 4ph scattering, $\Gamma^{4\mathrm{ph}}_{V_4V_4}$, and quintic-anharmonicity-assisted 3ph scattering, $\Gamma^{3\mathrm{ph}}_{V_3V_5}$, at 300~K. It can be seen the $V_3V_5$-3ph scattering is generally weaker than the ordinary $V_3V_3$-3ph channel, as expected from its higher perturbative order. However, its magnitude is comparable to that of the conventional $V_4V_4$-4ph scattering over a broad frequency range. This comparison directly confirms the diagrammatic ordering argument: although the $V_3V_5$ process involves a 3ph-like external topology, its contribution to the linewidth is of the same Van Hove order as 4ph scattering. A notable feature of Fig.~\ref{fig:diagram}(b) is that the frequency dependence of $\Gamma^{3\mathrm{ph}}_{V_3V_5}$ resembles that of $\Gamma^{3\mathrm{ph}}_{V_3V_3}$ more closely than that of $\Gamma^{4\mathrm{ph}}_{V_4V_4}$. This behavior follows from Eq.~\eqref{eq:g35}: the $V_3V_5$-3ph process has the same energy-conservation kernel and the same momentum-conservation condition as the conventional $V_3V_3$-3ph process. The main difference is the additional factor $(2n_3+1)$ that enhances the temperature dependence. As a result, the $V_3V_5$ channel may be viewed as a 3ph-like scattering process dressed by thermally populated higher-order anharmonic fluctuations.

The temperature enhancement of the $V_3V_5$-3ph scattering is evident at 1000~K, as shown in Fig.~\ref{fig:diagram}(c). Both $\Gamma^{3\mathrm{ph}}_{V_3V_5}$ and $\Gamma^{4\mathrm{ph}}_{V_4V_4}$ increase rapidly with temperature and become important corrections to the conventional 3ph linewidth. This behavior is consistent with their similar high-temperature scaling. In the classical limit, the ordinary $V_3V_3$-3ph linewidth scales approximately linearly with temperature ($\propto T$), whereas both the $V_4V_4$-4ph and the $V_3V_5$-3ph linewidths scale approximately as $T^2$. Importantly, the close magnitude of $\Gamma^{3\mathrm{ph}}_{V_3V_5}$ and $\Gamma^{4\mathrm{ph}}_{V_4V_4}$ in Si shows that the high-temperature correction to phonon lifetimes might not be attributed uniquely to the conventional 4ph scattering.

We next examine the effect of these scattering channels on $\kappa_l$. 
Fig.~\ref{fig:diagram}(d) compares $\kappa_l$ of Si calculated within the relaxation time approximation using different combinations of scattering rates. When only $V_3V_3$-3ph scattering is included, the calculated $\kappa_l$ follows a $T^{-1.09}$ dependence, consistent with the standard high-temperature behavior of 3ph-limited phonon transport. Including either the $V_4V_4$-4ph channel or the $V_3V_5$-3ph channel further suppresses $\kappa_l$ and accelerates its decay with temperature, thereby improving the agreement between simulation and experiment, particularly at elevated temperatures. Interestingly, the temperature dependences obtained upon including $V_4V_4$-4ph and $V_3V_5$-3ph scattering are similar, with fitted power laws close to $T^{-1.28}$ and $T^{-1.22}$, respectively. 
This similarity reflects the comparable scattering strengths and analogous temperature scaling of the two channels.

When both $V_4V_4$-4ph and $V_3V_5$-3ph scatterings are included in addition to $V_3V_3$-3ph scattering, the predicted $\kappa_l$ is further reduced and exhibits a slightly stronger temperature dependence, $\kappa_l \propto T^{-1.36}$. The resulting values are lower than the experimental measurements shown in Fig.~\ref{fig:diagram}(d). This quantitative underestimation may arise from several approximations employed in the present study. First, we adopted the relaxation time approximation to solve the Peierls-Boltzmann transport equation~\cite{Peierls1929}, without explicitly accounting for the collective behavior of phonons in Si~\cite{Cepellotti2016,albert2022hydrodynamic}. Second, the phonon wave-vector sampling mesh of $32^3$ remains insufficient to achieve fully converged absolute values of $\kappa_l$~\cite{guo2024sampling}. Third, the Perdew-Burke-Ernzerhof (PBE)~\cite{PBE} exchange-correlation~\cite{KohnXC} functional used here for Si has been shown to underestimate $\kappa_l$ in Si~\cite{jain2015effect}. 

Beyond these numerical approximations, such an underestimation may also point to possible missing physics in theoretical modeling. For example, only the quadratic contribution to the heat current is considered within the conventional Peierls-Boltzmann transport equation, whereas higher-order heat-current contributions may provide additional corrections to $\kappa_l$~\cite{sun2010lattice}, particularly at high temperatures. Moreover, the electronic contribution to thermal conductivity in Si due to bipolar effect may not be negligible at elevated temperatures~\cite{gu2020thermal}. We emphasize that the present calculations are designed to isolate the relative importance of $V_3V_5$-3ph scattering within a consistent framework, rather than to provide a fully optimized prediction of the experimental $\kappa_l$. The essential result is that $V_3V_5$-3ph scattering produces a contribution comparable to that of the conventional 4ph scattering.

The mode-resolved transport contributions further clarify the similarities and differences in the reduction of $\kappa_l$ caused by the additional scattering channels. Fig.~\ref{fig:diagram}(e) shows the cumulative $\kappa_l$ as a function of phonon frequency at 1000~K for Si. The dominant contribution to $\kappa_l$ arises from low- and intermediate-frequency acoustic phonons. Including either $V_4V_4$-4ph or $V_3V_5$-3ph scattering suppresses the cumulative $\kappa_l$ almost identically below 5~THz, whereas $V_4V_4$-4ph scattering leads to a slightly larger reduction in $\kappa_l$ for modes above 5~THz. This behavior is clearly reflected in the corresponding frequency-resolved phonon mean free paths. As shown in Figs.~\ref{fig:diagram}(f) and \ref{fig:diagram}(g), the reduction in the mean free paths of low-lying acoustic modes is very similar for the $V_4V_4$-4ph and $V_3V_5$-3ph processes, while the noticeably larger reduction in the mean free paths of high-lying optical modes ($\omega > 12$~THz) originates from the larger scattering rates associated with the $V_4V_4$-4ph process. These results show that the similarity between $V_4V_4$-4ph and $V_3V_5$-3ph scattering is not limited to the total $\kappa_l$, but also persists at the mode-resolved level. This close correspondence suggests that distinguishing the two channels experimentally may be challenging.

\begin{figure}[htp]
	\includegraphics[width = 1.0\linewidth]{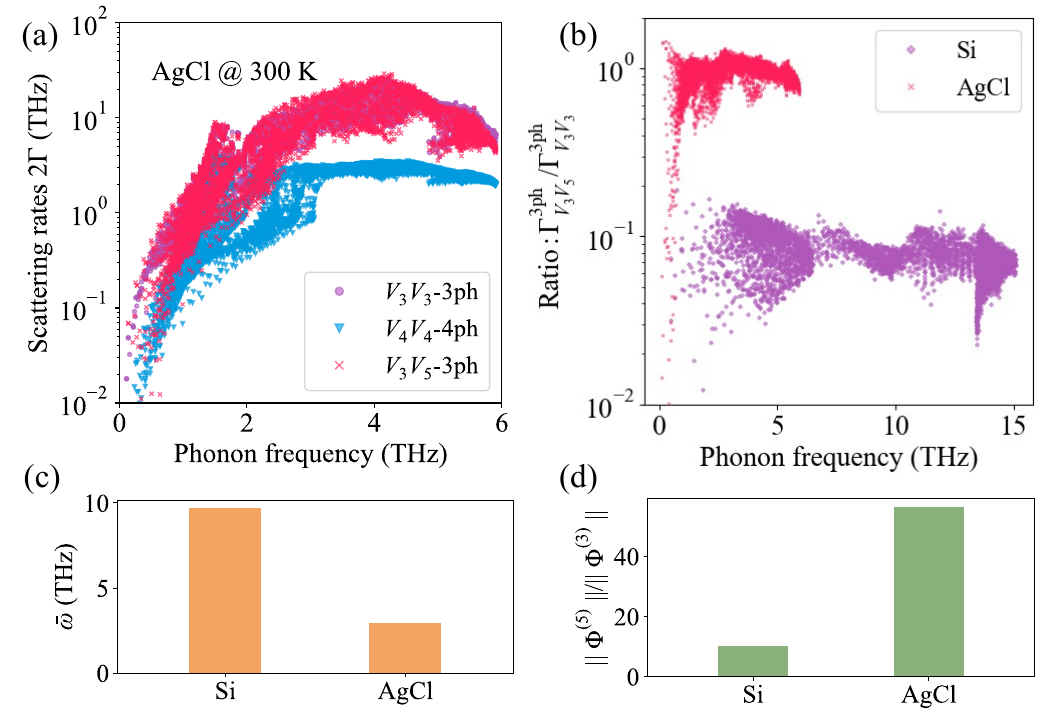}
	\caption{
    (a) Mode-resolved phonon scattering rates ($2\Gamma$) in AgCl at 300 K as a function of phonon frequency, comparing $V_3V_3$-3ph, $V_4V_4$-4ph, and $V_3V_5$-3ph channels.
    (b) Mode-resolved ratio $\Gamma^{\mathrm{3ph}}_{V_3V_5}/\Gamma^{\mathrm{3ph}}_{V_3V_3}$ for Si and AgCl.
    (c) Comparison of the mean phonon frequency $\bar{\omega}$ and the relative anharmonic force constant norm $\|\Phi^{(5)}\|/\|\Phi^{(3)}\|$ for Si and AgCl (see the $l^2$-norm of anharmonic constant in Appendix~\ref{sec:ifcs}).
}
	\label{fig:ratio}
\end{figure}

We finally examine AgCl to assess how the relative importance of the $V_3V_5$-3ph scattering changes in a more strongly anharmonic material. Fig.~\ref{fig:ratio}(a) compares the three scattering channels in AgCl at 300~K. In contrast to Si, the $V_3V_5$-3ph scattering rate in AgCl is not merely comparable to $V_4V_4$-4ph scattering; it approaches the magnitude of  $V_3V_3$-3ph scattering over a broad frequency range.  This result demonstrates that the quintic-anharmonicity-assisted 3ph scattering channel could become a leading contribution to phonon linewidths in materials with stronger higher-order anharmonicity.

To better understand the relative scattering strengths of the $V_3V_5$-3ph and $V_3V_3$-3ph processes, we show the mode-resolved ratio $\Gamma^{3\mathrm{ph}}_{V_3V_5} / \Gamma^{3\mathrm{ph}}_{V_3V_3}$ in Fig.~\ref{fig:ratio}(b). In Si, this ratio remains well below unity. In AgCl, however, the ratio is approximately one order of magnitude larger and can approach unity. This enhancement can be understood by noting that $\Gamma^{3\mathrm{ph}}_{V_3V_5}/\Gamma^{3\mathrm{ph}}_{V_3V_3} \propto (2n_3+1)V_5/V_3 \propto \omega_3^{-1} ||\Phi^{(5)}|| / ||\Phi^{(3)}||$, where $\Phi^{(n)}$ denotes the $n$th-order real-space interatomic force constants. Figs~\ref{fig:ratio}(c) and \ref{fig:ratio}(d) show the average phonon frequencies and the ratio between $||\Phi^{(5)}||$ and $||\Phi^{(3)}||$ in Si and AgCl, respectively, evaluated using the leading components. This approximate estimate gives $\bar{\omega}_{\rm Si} / \bar{\omega}_{\rm AgCl} = 3.3$ and $(||\Phi^{(5)}|| / ||\Phi^{(3)}||)_{\rm AgCl} / (||\Phi^{(5)}|| / ||\Phi^{(3)}||)_{\rm Si} = 5.5$, and therefore $(\Gamma^{3\mathrm{ph}}_{V_3V_5}/\Gamma^{3\mathrm{ph}}_{V_3V_3})_{\rm AgCl} / (\Gamma^{3\mathrm{ph}}_{V_3V_5}/\Gamma^{3\mathrm{ph}}_{V_3V_3})_{\rm Si} = 18.1$. This estimate recovers the same order of magnitude for  $\Gamma^{3\mathrm{ph}}_{V_3V_5}/\Gamma^{3\mathrm{ph}}_{V_3V_3}$ when comparing Si and AgCl as shown in Fig.~\ref{fig:ratio}(b).

Therefore, the enhanced relative strength of $\Gamma^{3\mathrm{ph}}_{V_3V_5}$ with respect to $\Gamma^{3\mathrm{ph}}_{V_3V_3}$ from Si to AgCl can be attributed to two factors. First, AgCl has lower characteristic phonon frequencies than Si, which increases the additional Bose occupation factor $(2n_3+1)$ at a given temperature. Second, the relative magnitude of the fifth-order anharmonic force constants with respect to the cubic force constants is larger in AgCl, indicating stronger higher-order anharmonicity. Together, these two effects explain why $\Gamma^{3\mathrm{ph}}_{V_3V_5}$ is much more important in AgCl than in Si. We note that we do not attempt to predict $\kappa_l$ in AgCl here, because our previous studies have shown that an accurate calculation of $\kappa_l$ in strongly anharmonic solids like AgCl requires the full frequency-resolved phonon spectral function beyond the quasiparticle approximation~\cite{xia2025lattice}. This treatment, which involves evaluating full frequency dependent phonon self-energies, is beyond the scope of the present study.



\textit{Conclusion.} We have identified and evaluated a previously overlooked phonon scattering channel arising from the coupled cubic-quintic anharmonic self-energy diagram. By classifying anharmonic phonon diagrams according to Van Hove's ordering parameter, we show that this quintic-anharmonicity-assisted three-phonon process is of the same perturbative order as the conventional four-phonon scattering, while retaining the energy-conservation structure of ordinary three-phonon scattering. First-principles calculations for Si show that this channel produces scattering rates comparable to those of four-phonon scattering and leads to a similar reduction and temperature dependence of the lattice thermal conductivity. This finding suggests that the high-temperature deviation from the conventional $T^{-1}$ behavior of lattice thermal conductivity, often attributed to four-phonon scattering, may also contain an important contribution from mixed-order scattering. The comparison between Si and AgCl further shows that the relative importance of this channel is strongly enhanced in materials with lower phonon frequencies and stronger higher-order anharmonicity, where it can approach the magnitude of ordinary three-phonon scattering. Our results therefore call for a reexamination of existing calculations through the explicit inclusion of mixed-order scattering, which may help reveal hidden microscopic thermal transport mechanisms by disentangling the underlying various kinds of anharmonic phonon scattering processes.

\begin{acknowledgments}
\textbf{Acknowledgments:} 
Y. X. acknowledges 1) the support from the U.S. National Science Foundation through awards No.~CBET-2445361 and No.~DMR-2317008, 2) the support from the Faculty Development Program at Portland State University, and 3) the computing resources provided by Bridges2 at Pittsburgh Supercomputing Center (PSC) through allocations mat220006p and mat220008p from the Advanced Cyber-infrastructure Coordination Ecosystem: Services \& Support (ACCESS) program, which is supported by National Science Foundation grants 2138259, 2138286, 2138307, 2137603, and 2138296. Y. X. is grateful to Z.J. W. for her encouragement and support during the preparation of this manuscript.
\end{acknowledgments}

\bibliography{CuSbS}

\clearpage
\onecolumngrid

\appendix

\section{Anharmonic force constants}\label{sec:ifcs}

\begin{figure}[!t]
    \centering
    \includegraphics[width=1.0\linewidth]{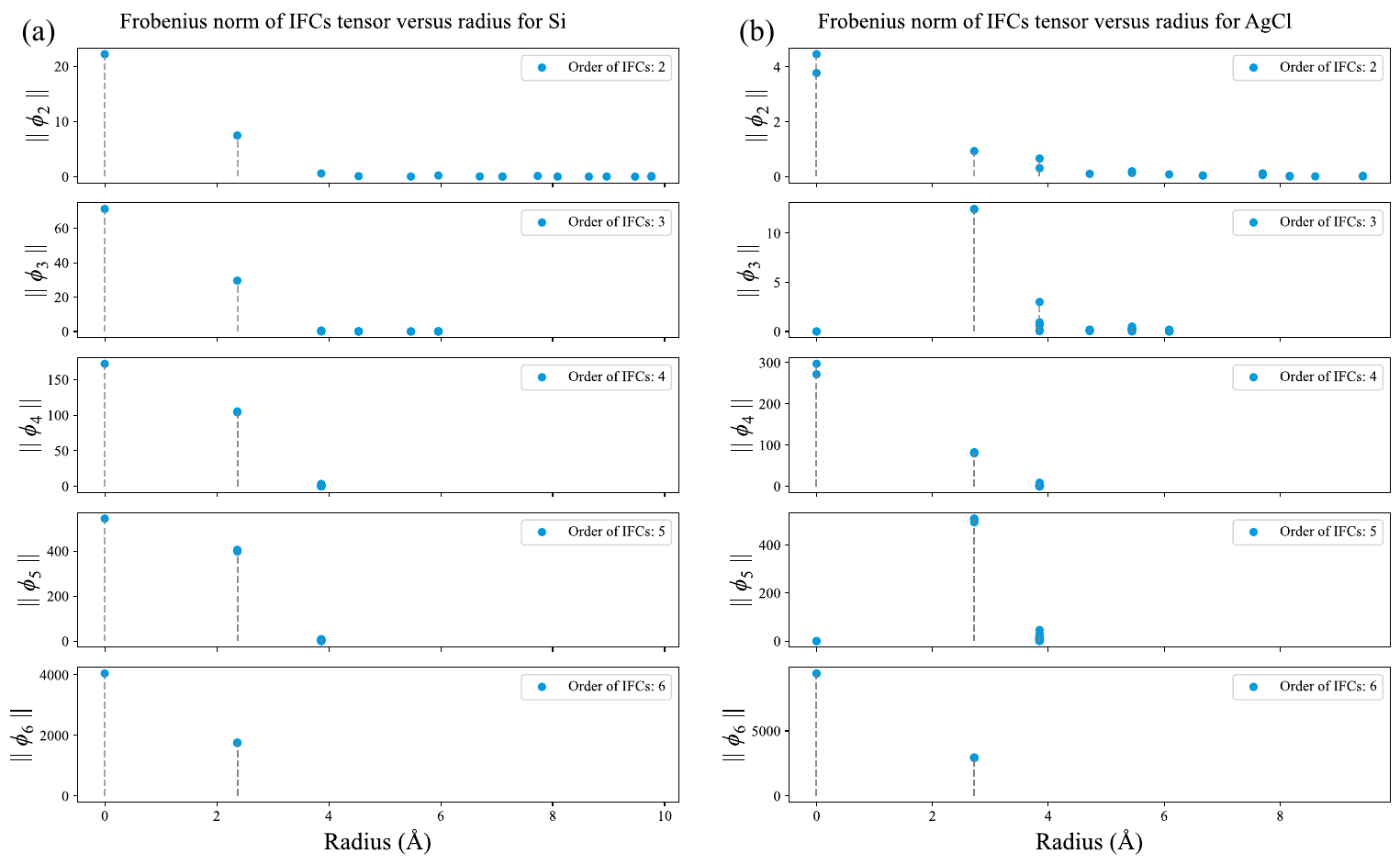}
    \caption{
    Frobenius norms of various orders of IFCs tensor versus radius for
    (a) silicon and (b) rocksalt AgCl.
    }
    \label{fig:ifcs_all}
\end{figure}

Anharmonic force constants up to sixth order were extracted using the Compressive Sensing Lattice Dynamics (CSLD) approach~\cite{csld,csldlong}. The first-principles calculation parameters and supercell sizes were adopted from our previous studies on silicon (Si)~\cite{xia2025first} and rocksalt AgCl~\cite{xia2025lattice}, whose convergence has been carefully examined. The obtained interatomic force constants (IFCs) of different orders are shown in Fig.~\ref{fig:ifcs_all}. To validate the IFCs extracted using CSLD, we further performed an independent calculation for Si using IFCs obtained from \textsc{hiphive} package~\cite{eriksson2019hiphive}, which yielded consistent phonon scattering rates.

\end{document}